\documentclass[aps,prd,reprint, superscriptaddress]{revtex4-1}
\usepackage{amsmath,amssymb, amsthm,amstext}
\usepackage{natbib}
\usepackage{graphicx}
\usepackage{color}
\usepackage{array, enumerate}
\usepackage{bm}
\usepackage{multirow}
\usepackage[breaklinks,colorlinks,citecolor=blue]{hyperref}
\usepackage{braket}
\usepackage{txfonts}

\def\mnras{MNRAS}

\def\apjl{Astrophys.J.Lett.}
\def\aj{Astronomical Journal}

\def\be{\begin{equation}}
\def\ee{\end{equation}}
\def\dotJI{\dot J_{\rm mig, I}}

\begin{document}
\title{Wet Extreme Mass Ratio Inspirals May Be More Common For \\
Spaceborne Gravitational Wave Detection}
\author{Zhen Pan}
\email{zpan@perimeterinstitute.ca}
\affiliation{Perimeter Institute for Theoretical Physics, Waterloo, Ontario N2L 2Y5, Canada}
\author{Zhenwei Lyu}
\affiliation{Perimeter Institute for Theoretical Physics, Waterloo, Ontario N2L 2Y5, Canada}
\affiliation{University of Guelph, Guelph, Ontario N1G 2W1, Canada}
\author{Huan Yang}
\email{hyang@perimeterinstitute.ca}
\affiliation{Perimeter Institute for Theoretical Physics, Waterloo, Ontario N2L 2Y5, Canada}
\affiliation{University of Guelph, Guelph, Ontario N1G 2W1, Canada}
\begin{abstract}
Extreme Mass Ratio Inspirals (EMRIs) can be classified as dry EMRIs and wet EMRIs based on their formation mechanisms.
Dry (or the ``loss-cone") EMRIs, previously considered as the main EMRI sources for the Laser Interferometer Space Antenna, are primarily produced by multi-body scattering in the nuclear star cluster and gravitational capture.
 In this work, we highlight an alternative EMRI formation channel: (wet) EMRI formation assisted
  by  the accretion flow around accreting galactic-center massive black holes (MBHs).
In this channel, the accretion disk captures stellar-mass black holes that are intially moving on inclined orbits, and subsequently drives them to  migrate towards the MBH - this process boosts the formation rate of EMRIs in such galaxies by orders of magnitude.
Taking into account  the fraction ($\mathcal O(10^{-2}-10^{-1})$)
of active galactic nuclei where the MBHs are expected to be rapidly accreting,
we forecast that wet EMRIs will contribute an important or even dominant fraction of all detectable EMRIs by spaceborne gravitational wave detectors.
\end{abstract}
\maketitle

\section{Introduction} The primary astrophysical sources for space-based gravitational wave detectors, such as   Laser Interferometer Space Antenna (LISA) \cite{LISA2019} and TianQin \cite{TianQin2020},
include massive black hole (MBH) bianaries, extreme mass ratio inspirals (EMRIs) \cite{Amaro-Seoane2007}, galactic binaries and stellar-mass black hole (sBH) binaries. Other systems, e.g. intermediate mass ratio inspirals \cite{Arca2019,Arca2020},
extremely large mass ratio inspirals \cite{Amaro2019,Amaro-Seoane2020} and cosmic strings \cite{Khakhaleva-Li2020}, may also be detectable, {\it albeit} with larger uncertainties. Among these sources, EMRIs provide unique opportunities in testing the Kerr spacetime \cite{Glampedakis2006, Barack2007}, probing the galactic-center cluster distribution \cite{Preto2010,Amaro2011,Pan2021}, understanding the astrophysical environmental effects \cite{Bonga:2019ycj,Yang:2019iqa,Barausse:2014tra}, and inferring the growth history of MBHs \cite{Berti:2008af,Gair2010,Pan:2020dze}.
 Loud EMRIs can serve as dark standard sirens for measuring the Hubble constant $H_0$ and the dark energy equation of state \cite{Laghi2021}.

EMRI formation mechanism can be classified into two main channels.
In the ``dry EMRI" channel, an EMRI may be produced after a sBH is gravitationally captured by a MBH,  following the
multi-body scatterings within the nuclear cluster
\cite{Amaro2018,Babak2017,Gair:2017ynp} (other processes involving tidal disruption or
tidal capture of binary sBHs, or tidal stripping of giant stars \cite{Miller2005,Chen2018,Wang2019,Raveh2021} may also contribute a fraction of dry EMRIs).
There are two characteristic
timescales \cite{Hopman2005,Bar-Or2016} in this process:
the GW emission timescale $t_{\rm gw}$ on which the sBH orbit shrinks
and the relaxation timescale $t_J$ on which the orbital angular momentum of the sBH changes, due to scatterings by stars and other sBHs. If $t_{\rm gw} > t_J$, the sBH will likely be randomly scattered either
into or away from the MBH (sBHs scattered into the MBH are known as prompt infalls).
If $t_{\rm gw} < t_J$, the sBH orbit
 gradually spirals into the MBH  to form an EMRI while random scatterings are negligible.
The generic rate can be obtained by solving the Fokker-Planck equation or
by N-body simulations \cite{Preto2010,Amaro2011,Pan2021}, subject to assumptions on the initial distributions of stars and sBHs in the nuclear cluster. In addition to the generic rate per MBH,
the EMRI rate density in the universe also depends on the mass function of MBHs, the fraction of MBHs living in
stellar cusps and the relative abundance of sBHs in stellar clusters. Taking into account these astrophysical
uncertainties, Babak et al. \cite{Babak2017,Gair:2017ynp} and Fan et al. \cite{Fan2020} forecasted that there will be a few to thousands of detectable (dry) EMRIs per year by LISA and TianQin, respectively.
In a recent paper \cite{Zwick2021}, Zwick et al. reanalyzed the GW emission timescales of inspiraling eccentric binaries
and realized Post-Newtonian (PN) corrections to the commonly used Peters' formula \cite{Peters1963} are necessary. With PN corrections implemented, the dry EMRI rate decreases by approximately at least one order of magnitude \cite{Vazquez2021}.

Wet EMRIs come from MBHs in gas-rich environments, where
the distributions of nearby stars and sBHs are significantly affected by the accretion flow.
About $1\%$ low-redshift ($z\lesssim 1$) galaxies and  $1\%-10\%$ high-redshift
 ($1\lesssim z\lesssim 3$) galaxies are active \cite{Galametz2009, Macuga2019} and known as active galactic nuclei (AGNs),
 in which galactic MBHs are believed to be rapdidly accreting gas in a disk configuration.
In the presence of an accretion disk, the periodic motion of a sBH generally generates density waves
which in turn affect the sBH's motion by damping both the orbital inclination with respect to the disk plane
and the orbital eccentricity, and driving the sBH's migration in the radial direction \cite{Goldreich1979,Goldreich1980,Tanaka2002,Tanaka2004}.
As long as the sBH is captured onto the disk, the density waves together with other disk-sBH interactions, e.g.,
head wind \cite{Yunes2011,Kocsis2011}, accelerate its inward migration until the vicinity of the MBH where GW emissions become prevalent. In addition to sBHs captured onto the disk,
star formation and subsequent birth of sBHs in the AGN disk
may also contribute to  wet EMRI formation \cite{Levin2003,Sigl2007,Levin2007}.
In this paper, we show that an accretion disk usually boosts the EMRI intrinsic rate per individual MBH by orders of magnitude compared with the loss-cone channel \footnote{More details can be found in a companion paper \cite{Pan2021}}. In particular, we suggest that wet EMRI formation is an important or even dominant channel for all observable EMRIs by spaceborne GW detectors.

The remaining part of this paper is organized as follows. In Sec.~\ref{sec:disk-sBH}, we summarize the interactions of
AGN disks with sBHs and stars. In Sec.~\ref{sec:FP}, we introduce the Fokker-Planck equation which governes the
evolution of sBHs and stars in a cluster with or without the presence of an AGN  disk.
In Sec.~\ref{sec:Gamma}, we present the generic dry EMRI rate per MBH and the wet EMRI rate per AGN.
In Sec.~\ref{sec:LISA}, we calculate the LISA detectable EMRI rate from both channels,
and we discuss the applications of wet EMRIs in Sec.~\ref{sec:app}.

Throughout this paper, we will use geometrical units $G=c=1$ and assume a flat
$\Lambda$CDM cosmology with $\Omega_{\rm m}=0.307,\Omega_{\rm \Lambda}=1-\Omega_{\rm m}$ and $H_0=67.7$ km/s/Mpc.

\section{ Disk-sBH and disk-star interactions}\label{sec:disk-sBH}
In addition to the gravitational forces from the MBH and the stars/sBHs in the cluster,
the orbital motion of a sBH around an accreting MBH is influenced by disk-sBH interactions:
density waves, head wind \cite{Goldreich1979,Goldreich1980,Tanaka2002,Tanaka2004,Kocsis2011} and other sub-dominant interactions including dynamic friction \cite{Chandrasekhar1943,Ostriker1999} and heating torque \cite{Masset2017,Hankla2020}.

As a sBH orbits around the MBH, its periodic motion excites density waves
consisting of three components \cite{Tanaka2002,Tanaka2004}: regular density waves arising from the circular motion, eccentricity waves arising from the non-circular motion and bending waves arising from the motion normal to the disk.
The density waves in turn affect the motion of the sBH:
the regular density waves exert a (type-I) migration torque on the sBH and drives its migration in the radial direction on the timescale $t_{\rm mig,I}$; the eccentricity and bending density waves damp the orbit eccentricity and the inclination with respect to the disk plane on the  timescale $t_{\rm wav}$.
Previous analytic studies \cite{Tanaka2002,Tanaka2004} calibrated with numerical simulations \cite{Paardekooper2010}
show that the type-I migration torque can be formulated as
\be\label{eq:JI}
\dotJI = C_{\rm I}\frac{m_{\rm bh}}{M} \frac{\Sigma}{M}\frac{r^4\Omega^2}{h^2}\ ,
\ee
where $m_{\rm bh}$ is sBH mass,
and $M=M(<r)$ is the total mass of the MBH, stars, sBHs and the disk within radius $r$;
the prefactor $C_{\rm I} = -0.85 + d\log \Sigma/d\log r + 0.9\ d\log T_{\rm mid}/d\log r$ depends on the disk profile;
$\Sigma(r), T_{\rm mid}(r), h(r), \Omega(r)$ are the disk  surface density,
the disk middle plane temperature, the disk aspect ratio and the sBH angular
velocity, respectively.
The corresponding migration timescale and damping timescale are
\be\label{eq:t2}
  t_{\rm mig,I} = \frac{J}{|\dotJI|}\sim
\frac{M}{m_{\rm bh}}\frac{M}{\Sigma r^2}\frac{h^2}{\Omega} , \quad
  t_{\rm wav} = \frac{M}{m_{\rm bh}}\frac{M}{\Sigma r^2}\frac{h^4}{\Omega}\ ,
\ee
where $J=r^2\Omega$ is the specific angular momentum of the sBH,
and $t_{\rm wav}\approx t_{\rm mig,I} h^2$, i.e., the eccentricity/inclination damping is
faster than the migration by a factor $h^2$. Therefore the orbit should become circular
long before the sBH migrate into the LISA band.
 A gap in the disk opens up if the sBH is so
massive that its tidal torque removes surrounding gas faster than the gas replenishment via viscous diffusion.
After a gap is opened, the type-I migration turns off and the sBH is subject to type-II migration driven
by a type-II migration torque $\dot J_{\rm mig,II}$ \cite{Syer1995}.

For a sBH embedded in the gas disk, surrounding gas in its gravitational
influence sphere flows towards it. Considering the differential rotation of the disk,
the inflow gas generally carries nonzero angular momentum relative to the sBH,
so that the inflow tends to circularize and form certain local disk or buldge profile around the sBH.
Depending on the radiation feedback and magnetic fields,
a major part of captured materials may escape in the form of outflow
and only the remaining part is accreted by the sBH \cite{Yang2014,McKinney2014}.
Because of the circularization process, it is reasonable to expect that the
outflow carries minimal net momentum with respect to the sBH.
As a result, the head wind in the influence sphere of the sBH is captured,
and the momentum carried by the wind eventually transfers to
the sBH. Therefore the specific torque exerted on the sBH from the
head wind is
\be\label{eq:Jwind}
\dot J_{\rm wind}^{\rm id} = - \frac{r \delta v_\phi \dot m_{\rm gas}}{m_{\rm bh}}\ ,
\ee
where the upper index ``id" denotes in-disk objects, $\delta v_\phi:=v_{\phi,\rm gas}- v_{\phi,\rm bh}$
is the head wind speed,
and $\dot m_{\rm gas}$ is the amount of gas captured per unit time (see \cite{Pan2021} for detailed calculation).

In summary,  the migration timescales of in-disk (id) sBHs and those outside (od) are
\be\label{eq:tmig}
t_{\rm mig}^{\rm bh,id} = \frac{J}{|\dot J_{\rm mig,I,II} + \dot J_{\rm gw} + \dot J_{\rm wind}|},\quad
t_{\rm mig}^{\rm bh,od} = \frac{J}{|\dot J_{\rm mig,I} + {\dot J_{\rm gw}}|} \ ,
\ee
where  $ \dot J_{\rm mig,I,II}= \dot J_{\rm mig,I} $ or $\dot J_{\rm mig,II}$
and $ \dot J_{\rm wind}= \dot J_{\rm wind}^{\rm id}$ [Eq.~(\ref{eq:Jwind})] or $0$,
depending on whether a gap is open. The specific torque arising from GW emissions is
\footnote{  GW emission turns out to has little influence on the wet EMRI rate which is determined by the capture and the migration of sBHs at large separations where GW emission is negligible.
We include the GW emission in the calculation because in some special cases
the migration torque $\dot J_{\rm mig, I}$ changes its sgin near the MBH
and the GW emission works to overcome the would-be migration trap \cite{Pan2021}. }
\be\label{eq:Jgw}
\dot J_{\rm gw} = -\frac{32}{5} \frac{m_{\rm bh}}{M}\left(\frac{M}{r}\right)^{7/2}\ .
\ee
The damping timescale of sBH orbital inclination and eccentricity
is given by Eq.~(\ref{eq:t2})
\be
t_{\rm wav}^{\rm bh,od} = \frac{M}{m_{\rm bh}}\frac{M}{\Sigma r^2}\frac{h^4}{\Omega}\ .
\ee
The above discussion of disk-sBH interactions also equally applies to stars in the cluster,
except stars are usually lighter ($m_{\rm star} < m_{\rm bh}$),
and the head wind impact on stars is weak ($\dot J_{\rm wind}^{\rm star}\approx 0$) considering that
the wind could be largely suppressed in the presence of star radiation feedback and solar wind \cite{Gruzinov2020,Li2020}.
Becuse the structure of AGN disks has not been fully understood, we consider
three commonly used AGN disk models: $\alpha$-disk, $\beta$-disk \cite{Sirko2003} and TQM disk \cite{Thompson2005} in this work.

\section{Fokker-Planck equation}\label{sec:FP}
Statistical properties of stars and sBHs in the stellar cluster
are encoded in their distribution functions $f_i(t, E, R)$ ($i={\rm star/bh}$) in the phase space, where
\be
E:=\phi(r)-v^2(r)/2\ , \quad  R:=J^2/J_c^2(E)
\ee
are the specific orbital (binding) energy and the normalized orbital angular momentum, respectively.
Here $\phi(r)$ is the (positive) gravitational potential, $v$ is the orbital speed,
and $J_c(E)$ is the specific angular momentum of
a circular orbiter with energy $E$. Given initial distributions $f_i(t=0, E, R)$, the subsequent
evolution is governed by the orbit-averaged Fokker-Planck equation.
In the case of no gas disk, the Fokker-Planck equation (for both stars and sBHs)
is formulated as \cite{Cohn1978,Cohn1979,Binney1987}
\be\label{eq:FP}
  \mathcal C\frac{\partial f}{\partial t}
  = - \frac{\partial}{\partial E} F_E
  - \frac{\partial}{\partial R}F_R \ ,
\ee
where $f=f_i(t,E,R)$, $\mathcal C=\mathcal C(E,R)$ is a normalization coefficient, and
$F_{E,R}$  is the flux in the $E/R$ direction:
\be\label{eq:flux}
\begin{aligned}
  -F_E &= \mathcal D_{EE}\frac{\partial f}{\partial E} + \mathcal D_{ER}\frac{\partial f}{\partial R} + \mathcal D_E f\ ,\\
  -F_R &= \mathcal D_{RR}\frac{\partial f}{\partial R} + \mathcal D_{ER}\frac{\partial f}{\partial E} + \mathcal D_R f\ ,
\end{aligned}
\ee
where the diffusion coefficients $\{\mathcal D_{EE}, \mathcal D_{ER}, \mathcal D_{RR}\}_i$ and the
advection coefficients $\{\mathcal D_E, \mathcal D_R\}_i$ are functions of $f_i(t, E, R)$ \cite{Cohn1978,Cohn1979,Binney1987}.
From flux $\{F_E, F_R\}_{\rm bh}$, we can compute the EMRI rate via the lose cone mechanism as
\be\label{eq:emri}
\Gamma_{\rm lc}(t) = \int_{E>E_{\rm gw}}  \vec F \cdot d\vec l\ ,
\ee
where $\vec F = (F_E, F_R)$, $d\vec l = (dE, dR)$ is the line element along the boundary of the loss cone,
and $E_{\rm gw}$ is a characteristic energy scale above which the sBH
GW emission is dominant with $t_{\rm gw}<t_J$ \cite{Hopman2005, Amaro2011, Preto2010,Amaro2011,Pan2021, Zwick2021, Vazquez2021}.

In the presence of an AGN disk, stars and sBHs settle as two components: a cluster component and a disk component.
We expect the distribution functions of cluster-component stars and sBHs
acquire some dependence on the orbital inclination as interacting with the disk. For convenience,
we choose to integrate out the inclination  and work with the inclination-integrated
 distribution functions $f_i(t,E,R)$ of the cluster-component stars and sBHs.
Considering the density waves excited on the disk  to
damp the orbital inclinations and eccentricities of orbiters,
and   to drive the orbiters' inward migration together with head winds and GW emissions,
we rewrite the Fokker-Planck equation as
\be\label{eq:FP_disk}
  \mathcal C\frac{\partial f}{\partial t}
  = - \frac{\partial}{\partial E} F_E
  - \frac{\partial}{\partial R}F_R + S\ ,
\ee
where flux $F_{E,R}$ are defined in Eq.~(\ref{eq:flux}), with the advection coefficients modified by
disk-star/sBH interactions as
\be\label{eq:coef}
\begin{aligned}
  \mathcal  D_{E,\rm bh} &\rightarrow \mathcal D_{E,\rm bh}  - \mathcal C\frac{E}{t_{\rm mig}^{\rm bh, od}}\ ,\quad
  &&\mathcal D_{R,\rm bh} \rightarrow \mathcal D_{R,\rm bh}  - \mathcal C\frac{1-R}{t_{\rm wav}^{\rm bh, od}}\ , \\
  \mathcal D_{E,{\rm star}} &\rightarrow \mathcal D_{E,{\rm star}}  -\mathcal C\frac{E}{t_{\rm mig}^{\rm star, od}}\ ,\quad
  &&\mathcal D_{R,{\rm star}}\rightarrow  \mathcal D_{R,{\rm star}} - \mathcal C\frac{1-R}{t_{\rm wav}^{\rm star, od}}\ ,\nonumber
\end{aligned}
\ee
and the negative source term $S=S_i(t,E,R)$ arising from spherical-component stars/sBHs captured onto the disk
is parameterized as
\be\label{eq:source}
S_{\rm bh} = -\mu_{\rm cap}\mathcal C \frac{f_{\rm bh}}{t_{\rm mig}^{\rm star, id}}\ ,\quad
S_{\rm star} = -\mu_{\rm cap}\frac{m_{\rm star}}{m_{\rm bh}}\mathcal C  \frac{f_{\rm star}}{t_{\rm mig}^{\rm star, id}}\ ,
\ee
with $\mu_{\rm cap}\in [h,1]\frac{m_{\rm bh}}{m_{\rm star}}$ a phenomenological parameter quantifying the
disk capture efficiency (see \cite{Pan2021} for more details).
A new EMRI forms if a sBH is captured onto the disk and migrate
to the vicinity of the MBH within the disk lifetime $T_{\rm disk}$,
therefore the EMRI rate assisted by the AGN disk is given by
\be\label{eq:emri_disk}
\Gamma_{\rm disk}(t;T_{\rm disk}) = \int\int_{ t_{\rm mig}^{\rm bh,id} < T_{\rm disk}} -S_{\rm bh}(t, E,R)\ dEdR\ .
\ee

\section{EMRI rate per MBH/AGN: dry and wet}\label{sec:Gamma}

\subsection{Dry EMRIs}
Given initial distributions of stars and sBHs in the stellar cluster, one can evolve
the system according to the Fokker-Planck equation (\ref{eq:FP}) and calculate the EMRI rate
in the loss cone channel using Eq.~(\ref{eq:emri}).
As shown in Refs.~\cite{Preto2010,Amaro2011,Pan2021}, the EMRI rate mainly depends on the total number of
stars within the MBH influence radius, which determines the relaxation timescale and the relative abundance of sBHs
in the stellar cluster. Following Ref.~\cite{Babak2017}, the time-averaged EMRI rate per MBH can be parameterized as
\be\label{eq:dry_eff}
{\Gamma_{\rm dry}}(M_\bullet; N_p) = C_{\rm dep}(M_\bullet; N_{p})
C_{\rm grow}(M_\bullet; N_{p})\Gamma_{\rm lc}(M_\bullet)\ ,
\ee
with
\be\label{eq:dry}
\Gamma_{\rm lc}(M_\bullet) = 30 \left( \frac{M_\bullet}{10^6 M_\odot}\right)^{-0.19}\ {\rm Gyr}^{-1}\ ,
\ee
where $N_p$ is the average number of prompt infalls per EMRI;
$C_{\rm dep} $ and $C_{\rm grow}$ are  correction factors accounting for possible depletion of sBHs in
the cusp as sBHs accreted by the MBH and capping the maximum MBH growth
via accreting sBHs, respectively, and the loss-cone EMRI rate in Eq.~(\ref{eq:dry}) is lower than previous calculations
\cite{Preto2010,Amaro2011, Babak2017,Pan2021} by one order of magnitude because
these previous results were based on the Peters' formula \cite{Peters1963}
which underestimate the GW emission timescales
of eccentric binaries and the true EMRI rate should be
lower by approximately at least one order of magnitude \cite{Zwick2021,Vazquez2021}.

Following Ref.~\cite{Babak2017}, we explain the two corrections $C_{\rm dep}(M_\bullet;N_p)$ and $C_{\rm grow}(M_\bullet;N_p)$  to the generic dry EMRI rate.
Consider a MBH with mass $M_\bullet$,  whose influence sphere ($r<r_c=2M_\bullet/\sigma^2$)
 encloses a number of sBHs with total mass $\Sigma m_{\rm bh} \simeq 0.06 M_\bullet$,
 and these sBHs will be depleted by the MBH via EMRIs and prompt infalls on a timescale
\be
\begin{aligned}
  T_{\rm dep}(r_c)
  &= \frac{\sum m_{\rm bh}}{(1+N_p) \Gamma_{\rm lc}(M_\bullet) m_{\rm bh}}\\
  &=\frac{200}{1+N_p}\left(\frac{m_{\rm bh}}{10 M_\odot}\right)^{-1}
  \left(\frac{M_\bullet}{10^6M_\odot}\right)^{1.19}\ {\rm Gyr}\ ,
\end{aligned}
\ee
where $N_p$ is the average number of prompt infalls per EMRI.
On the influence sphere, the relaxation timescale of the star cluster is approximately \cite{Binney1987}
\be
T_{\rm rlx}(r_c) \simeq \left( \frac{\sigma}{20 {\rm km/s}} \right) \left(\frac{r_c}{1 {\rm pc}}\right)^2\ {\rm Gyr}\ ,
\ee
where the velocity dispersion is related to the MBH mass by the famous $M_\bullet-\sigma$ relation \cite{Gultekin2009}.
The depletion correction $C_{\rm dep}$ is defined as
\be
C_{\rm dep}(M_\bullet;N_p) := {\rm min.}\left\{\frac{T_{\rm dep}}{T_{\rm rlx}}, 1\right\}\ ,
\ee
where
\be
\frac{T_{\rm dep}}{T_{\rm rlx}}\simeq \frac{12}{1+N_p}\left(\frac{m_{\rm bh}}{10 M_\odot}\right)^{-1}
\left(\frac{M_\bullet}{10^6M_\odot}\right)^{0.06}\ .
\ee

\begin{figure*}
\includegraphics[scale=0.6]{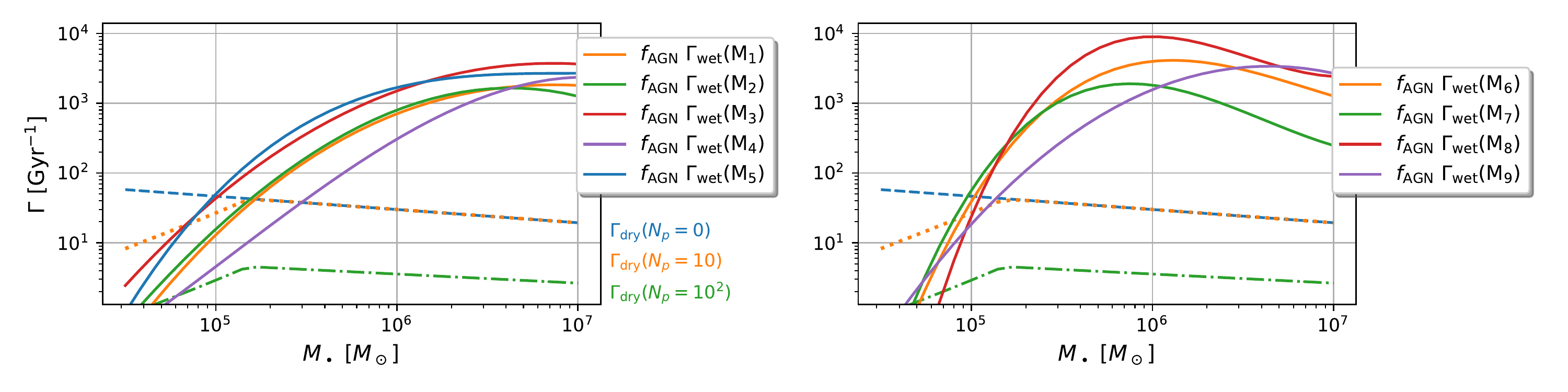}
\caption{\label{fig:Gamma} Average EMRI rates per MBH in the loss cone channel
$\Gamma_{\rm dry}(M_\bullet; N_p)$ and per AGN in the disk channel
$\Gamma_{\rm wet}(M_\bullet;\mathbb{M})$, where $N_p$ is the number
of prompt infalls per EMRI, and $\mathbb{M}$ consists of all model parameters
of initial condition of stellar clusters,
AGN duty cycles and AGN disk model, where the AGN fraction is $f_{\rm AGN}=1\%$.}
\end{figure*}

The growth correction
\be
C_{\rm grow} := {\rm min.}\left\{e^{-1} \frac{M_\bullet}{\Delta M_\bullet},1\right\}
\ee
arises from requiring the MBH grows no more than $e^{-1}$ via accreting sBHs, where
\be
\Delta M_\bullet = m_{\rm bh} (1+N_p) C_{\rm dep}(M_\bullet;N_p) \Gamma_{\rm lc}(M_\bullet) T_{\rm emri}(M_\bullet)\ ,
\ee
is the MBH growth via accreting sBHs, and
\be
T_{\rm emri}(M_\bullet) = \int dz \frac{dt}{dz} C_{\rm cusp}(M_\bullet, z)
\ee
is the effective growth time when the MBH lives in a
stellar cusp.

In Fig.~\ref{fig:Gamma}, we show 3 sample models of dry EMRIs with $N_p=\{0,10,10^2\}$,
where $\Gamma_{\rm dry}(N_p=0)$ is the same as the generic rate
[Eq.~(\ref{eq:dry})] in the mass range of interest, $\Gamma_{\rm dry}(N_p=10)$ is capped by
the accretion growth limit $C_{\rm grow}$ for light MBHs, and $\Gamma_{\rm dry}(N_p=10^2)$ is further
 reduced by the sBH depletion $C_{\rm dep}$ across the entire mass range.

\subsection{Wet EMRIs}
More technical complications are involved in calculating the wet EMRI rate due to the uncertainties
in AGN accretion history, AGN accretion disks and initial conditions of stellar clusters, which we outline as follows in accordance with our previous work \cite{Pan2021}.

In this paper, we conservatively assume a constant AGN fraction $f_{\rm AGN}=1\%$ throughout the universe,
though it can be  $10$ times higher \cite{Galametz2009,Macuga2019}. Being consistent with the AGN fraction, the total duration
of active phases of an AGN is about $10^8$ yr \cite{Soltan1982}, therefore MBHs are in quiet phase most of the time.
Another complication is that AGN accretion is likely episodic \cite{King2015,Schawinski2015}, i.e., a MBH may become active
for multiple times during its whole life.
Without detailed knowledge of the duty cycle of an MBH, we simplify it as
a long quiet phase of $T_0=5$ Gyr followed by a short active phase
of $T_{\rm disk}=10^7$ or $10^8$ yr. For $T_{\rm disk}=10^8$ yr, there is only one active phase.
On the other hand, there are on average 10 active phases for models with $T_{\rm disk}=10^7$ yr,
and we only consider the following two extremal cases. If a (low-redshift) AGN that has gone through all the 10 active phases and the relaxation between different
active phases is not expected to substantially change the sBH distribution, the average EMRI rate is approximately
same to that in the case of $T_{\rm disk}=10^8$ yr.
If a (high-redshift) AGN that has gone through only 1 active phase, the duty cycle is simply
a long quiet phase with duration $T_0=5$ Gyr followed by a short active phase with duration $T_{\rm disk}=10^7$ yr.
That is to say, $T_{\rm disk}$ in our model is approximately the total duration of \emph{all} active phases an AGN has gone through.

The structure of AGN disks has not been fully understood either, partially due  to
the large range over which an AGN disk extends: from an inner radius of a few gravitational radii of the MBH to the
outer radius of parsec scale where the AGN disk connects to the galactic gas disk.
Three commonly used AGN disk models: $\alpha$-disk, $\beta$-disk \cite{Sirko2003} and TQM disk \cite{Thompson2005},
are different in their prescriptions of disk viscosity and/or disk heating mechanism which lead to large differences
in predicted disk structures.
Each disk model is specified by two model parameters, the MBH accretion rate $\dot M_\bullet$
and a viscosity parameter \cite{Pan2021}:
an $\alpha$ parameter which prescribes the ratio between the viscous stress and the local
total/gas pressure in the $\alpha/\beta$-disk and a $X$ parameter
which prescribes the ratio between the radial gas velocity
and the local sound speed in the TQM disk. For calculating the wet EMRI rate, we consider both
an $\alpha$-disk model \cite{Sirko2003} with the viscosity parameter $\alpha=0.1$ and
accretion rate $\dot M_\bullet=0.1 \dot M_\bullet^{\rm Edd}$,
and a TQM disk \cite{Thompson2005} with the viscosity parameter $X=0.1$
and accretion rate $\dot M_\bullet=0.1 \dot M_\bullet^{\rm Edd}$
($\beta$-disk is different from $\alpha$-disk only in the inner region where
radiation pressure dominates over gas pressure, and this difference has little impact on the wet EMRI rate).

For calculating the wet EMRI rate, we also need to specify the initial distributions of stars and sBHs in the stellar cluster,
which we assume the commonly used Tremaine's cluster model \cite{Tremaine1994} with a power-law  density
profile $n_{\rm star}(r)\sim r^{-\gamma}$ deep inside the influence sphere
of the MBH and $n_{\rm star}(r)\sim r^{-4}$ far outside,
and sBHs are of the same density profile with a relative abundance $\delta$.

Given initial distributions of stars and sBHs in the stellar cluster,
we first evolve the system for time $T_0$ according to the Fokker-Planck equation (\ref{eq:FP}),
then turn on an accretion disk and continue the evolution for time $T_{\rm disk}$, according to the modified
Fokker-Planck equation (\ref{eq:FP_disk}). In the active phase, the disk assisted EMRI rate
is computed using Eq.~(\ref{eq:emri_disk}). We show  the time-averaged EMRI rate per AGN
\be\label{eq:Gamma_disk}
\Gamma_{\rm wet}(M_\bullet; \mathbb{M})
= \frac{1}{T_{\rm disk}}\int_{T_0}^{T_0+T_{\rm disk}}\Gamma_{\rm disk}(t, M_\bullet;\mathbb{M}) dt\ ,
\ee
for different models  $\mathbb{M}$ in Fig.~\ref{fig:Gamma}, where $\mathbb{M}$ denotes models
parameterizing initial distributions of stars and sBHs in the cluster,
duty cycles of MBHs and AGN disk model (see Table~\ref{table} for model parameters for
all the $9$ models considered in this work).

Because sBHs are captured onto the disk and migrate inward efficiently,
and the sBH loss via prompt infalls is negligible ($N_p \ll 1$),
the wet EMRI rate is mainly limited by the number of sBHs available in the stellar cluster.
As a result, we find the presence of an AGN disk usually boosts the EMRI formation
rate by orders of magnitude \cite{Pan2021,Tagawa2020} regardless of the variations of
different disk models considered.

\section{ Total and LISA detectable EMRI rates}\label{sec:LISA}
For calculating the total EMRI rate, we consider two redshift-independent
MBH mass functions in the range of $(10^4, 10^7) M_\odot$,
\be\label{eq:fbullet}
\begin{aligned}
  f_{\bullet,-0.3}: \frac{dN_\bullet}{d \log M_\bullet}
  &= 0.01 \left( \frac{M_\bullet}{3\times 10^6 M_\odot}\right)^{-0.3}\ {\rm Mpc}^{-3}\ ,\\
    f_{\bullet,+0.3}: \frac{dN_\bullet}{d \log M_\bullet}
  &= 0.002 \left( \frac{M_\bullet}{3\times 10^6 M_\odot}\right)^{+0.3}\ {\rm Mpc}^{-3}\ ,
\end{aligned}
\ee
where the former one is approximate to the mass function as modelled in Refs.~\cite{Barausse2012,Sesana2014,Antonini2015,Antonini2015b} assuming
MBHs were seeded by Population III stars and accumulated mass via mergers
and gas accretion along cosmic history, and the latter one is a phenomenological model \cite{Gair2010}.
The differential EMRI rates (in observer's frame) in the two formation channels are written as
\be\label{eq:R2}
\begin{aligned}
  \frac{d^2\mathcal R_{\rm dry}}{dM_\bullet dz}
  &=\frac{1}{1+z} \frac{dN_\bullet}{dM_\bullet}\frac{dV_{\rm c}(z)}{dz} C_{\rm cusp}(M_\bullet,z)
  \Gamma_{\rm dry}(M_\bullet; N_{p})\ , \\
  \frac{d^2\mathcal R_{\rm wet}}{dM_\bullet dz}
  &= \frac{f_{\rm AGN}}{1+z}\frac{dN_\bullet}{dM_\bullet}\frac{dV_{\rm c}(z)}{dz}C_{\rm cusp}(M_\bullet,z)
  \Gamma_{\rm wet}(M_\bullet;\mathbb{M})\ ,
\end{aligned}
\ee
where the factor $1/(1+z)$ arises from the cosmological redshift,
$V_{\rm c}(z)$ is the comoving volume of the universe up to redshift $z$,
$C_{\rm cusp}(M_\bullet,z)$ is the fraction of MBHs living in stellar cusps
which are supposed to be evacuated during mergers of binary MBHs and re-grow afterwards \cite{Barausse2012,Sesana2014,Antonini2015,Antonini2015b}.
For cases with mass function $f_{\bullet,+0.3}$, we use the same $C_{\rm cusp}$ function as in \cite{Babak2017}
and we simply take $C_{\rm cusp}=1$ for cases with phenomenological mass function $f_{\bullet,+0.3}$
\footnote{There was some misplot of the $C_{\rm cusp}(M_\bullet,z)$ function in \cite{Babak2017},
and we thank Alberto Sesana for kindly providing the correct one.}.

\begin{figure*}
\includegraphics[scale=0.24]{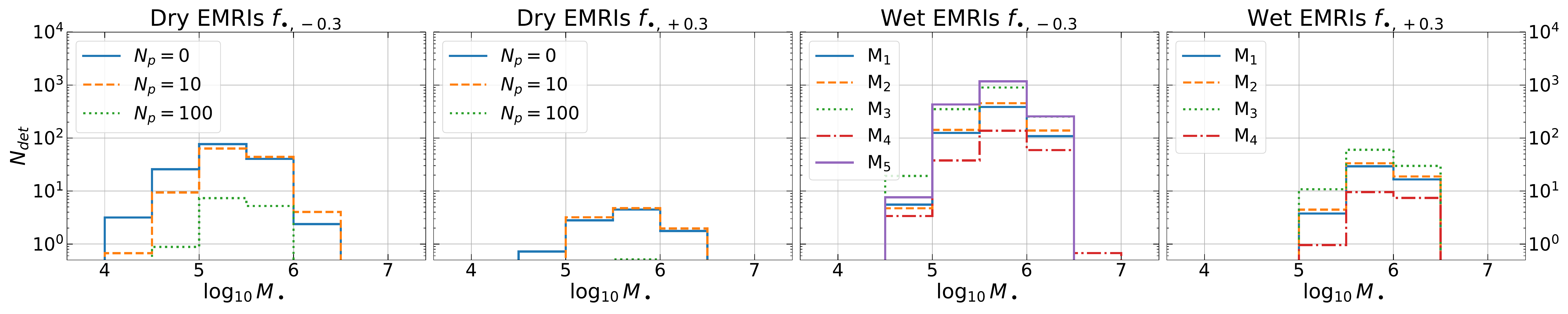}
\caption{\label{fig:Ndet} Forecasted LISA detectable dry and wet EMRI rates $N_{\rm det}$ per mass bin
($M_\bullet [M_\odot]$) per year
for different models, where $f_{\bullet,\pm0.3}$ are the two different MBH mass functions [Eq.~(\ref{eq:fbullet})],
$N_p$ is the number of prompt infalls per EMRI in the loss cone channel,
and the wet EMRI model parameters of M$_{1,...,5}$
are detailed in Table~\ref{table}.}
\end{figure*}

In order to calculate the LISA detectable EMRI rate in each channel, we
construct a population of EMRIs with sBH mass $m_{\rm bh}=10M_\odot$, MBH spin $a=0.98$,
and MBH masses and redshifts randomly sampled according to the differential EMRI rates [Eq.~(\ref{eq:R2})].
For each individual EMRI, we need 10 more parameters to uniquely specify
its binary configuration at coalescence and its gravitational waveform
\cite{Barack:2003fp,Chua:2015mua,Chua:2017ujo}:
sky localization $\hat{\bm{n}}$,
MBH spin direction $\hat{\bm{a}}$, 3 phase angles,
coalescence time $t_0$, inclination angle $\iota_0$ and eccentricity $e_0$
at coalescence.
For both dry and wet EMRIs, we assume the sky locations and
the MBH spin directions are isotropically distributed on the
sphere, 3 phase angles are uniformly distributed in $[0, 2\pi]$, coalescence times are randomly sampled from $[0, 2]$ yr,
and cosines of inclination angles are randomly sampled from $[-1,1]$. Distributions of eccentricity $e_0$ are different: uniform disbribution of $e_0$ in $[0, 0.2]$ for dry EMRIs v.s. $e_0=0$ for wet EMRIs.

For each EMRI, we compute its time-domain waveform $h_{+,\times}(t)$ using the Augment
Analytic Kludge (AAK) \cite{Barack:2003fp,Chua:2015mua,Chua:2017ujo} with the conservative Schwarzschild plunge condition,
because the PN corrections used for constructing the AAK waveform model are increasingly inaccurate as the orbital separation decreases. Extending the waveform to the Kerr last stable orbit likely leads to an
overestimate of the signal-to-noise ratio (SNR) \cite{Babak2017,Chua:2017ujo}.
The SNR is calculated as a noise weighted inner product in the frequency domain \cite{Finn1992}
\be
{\rm SNR} = \sqrt{4\int_0^\infty \frac{h_+(f)h_+^*(f)+h_\times(f)h_\times^*(f)}{S_{\rm n}(f)}\ df}\ ,
\ee
where $S_{\rm n}(f)$ is the sky-averaged detector sensitivity of LISA \cite{LISA2017,Babak2017}.
The expected LISA detectable EMRI rates (SNR$\geq20$) of different models in each mass bin are shown in Fig.~\ref{fig:Ndet},
and the total event rates and the LISA detectable rates are collected in Table~\ref{table}.
From Fig.~\ref{fig:Ndet} and Table~\ref{table},
wet EMRI formation is evidently an important or even dominant channel for all the models we have considered.

\begin{table*}
\caption{\label{table} Comparison of dry and wet EMRI rates in different models, where $f_\bullet$ is the MBH mass function.
The last two columns are the total EMRI rate in the redshift range of $0<z<4.5$
and the corresponding LISA detectable (SNR$\geq 20$) rate.}
\resizebox{2.1\columnwidth}{!}{
\addtolength{\tabcolsep}{5pt}
\begin{tabular}{c c c@{\hskip 0.5cm}*{4}{c}  r r }
      Dry EMRIs &  $f_\bullet$ &  $N_p$ & & & & & Total rate [yr$^{-1}$] & LISA detectable rate [yr$^{-1}$]\\
\cline{1-9}
 & $f_{\bullet,-0.3}$  &  $0$ &&&&& 3500  & 150  \\
                     && $10$   &&&&& 1300 & 120  \\
                     && $10^2$ &&&&& 150 & 14   \\
                     &$f_{\bullet,+0.3}$ &  $0$ &&&&& 160 & 10 \\
                     &  &$10$    &&&&& 130 & 10 \\
                     &  & $10^2$ &&&&& 15 & 1 \\

\hline
 Wet EMRIs &$f_\bullet$ & $\mathbb{M}:$ & ($\gamma, \delta$) & $\mu_{\rm cap}$ &  $(T_{\rm disk}\ [{\rm yr}], f_{\rm AGN})$ & AGN Disk
 & Total rate [yr$^{-1}$] & LISA detectable rate [yr$^{-1}$]\\\cline{2-9}
     & $f_{\bullet,-0.3}$ & ${\rm M}_1:$ & (1.5, 0.001) & 1
&$(10^8, 1\%)$  & $\alpha$-disk & 11000 & 600\\
             &  & ${\rm M}_2:$ &(1.5, 0.001) & 0.1 &  & & 11000 & 760\\
             &  & ${\rm M}_3:$& (1.5, 0.002) & 1   &  & & 24000 & 1500\\
             &  & ${\rm M}_4:$& (1.8, 0.001) & 1   &  & & 8100  & 240\\
             &  & ${\rm M}_5:$& (1.5, 0.001) & 1   & $(10^8, 1\%)$ &TQM disk &23000 & 1900\\
             &  & ${\rm M}_6:$ &(1.5, 0.001) & 1   & $(10^7, 1\%)$ & $\alpha$-disk &39000 &4200\\
             &  & ${\rm M}_7:$& (1.5, 0.001) & 0.1 & &  & 21000 & 3000\\
             &  & ${\rm M}_8:$& (1.5, 0.002) &  1  & &  & 80000 & 9800\\
             &  & ${\rm M}_9:$& (1.8, 0.001) & 1   & &  & 22000 & 1400\\
             & $f_{\bullet,+0.3}$ & ${\rm M}_{1}:$
             & (1.5, 0.001) & 1 & $(10^8, 1\%)$ & $\alpha$-disk & 2100 & 49\\
             & &${\rm M}_{2}:$ & (1.5, 0.001) & 0.1 & & & 2000 & 57 \\
             & &${\rm M}_{3}:$ & (1.5, 0.002) & 1   & & & 4300 & 100 \\
             &  &${\rm M}_{4}:$& (1.8, 0.001) & 1   & & & 1900 & 18\\
            \hline
            \hline
\end{tabular}
}
\end{table*}

\section{ Applications of wet EMRIs}\label{sec:app}
Due to the high LISA sensitivity to the EMRI eccentricity whose value at coalescence can be
measured with typical uncertainty as low as $10^{-5}$ \cite{Babak2017},
wet EMRIs can be distinguished from dry ones via eccentricity measurements,
as wet EMRIs are expected to be circular in the LISA band as a result of the efficient eccentricity damping
by the density waves ($t_{\rm wav}\ll t_{\rm mig}$), while
dry EMRIs from the loss-cone channel
are highly eccentric as entering the LISA band and remain
mildly eccentric at coalescence \cite{Hopman2005,Babak2017}.
Another subdominant dry EMRI channel involving tidal stripping of giant stars
seems unlikely to produce such circular EMRIs either \cite{Wang2019}, while the prediction of the channel involving tidal disruption of binary sBHs is more uncertain \cite{Miller2005,Raveh2021}.  The disk-environmental effects may produce measurable phase shift in the EMRI waveform  \cite{Kocsis2011,Yunes2011,Barausse2014}.

EMRIs have unprecedented potential  to probe fundamental laws of gravity and the nature of dark matter  \cite{Glampedakis2006,Barack2007,Hannuksela:2018izj,Zhang:2019eid}.
In previous studies, such tests using EMRIs have been implicitly
assumed in vacuum without any environmental contamination. However, as we have shown here,
wet EMRIs are possibly more common in the universe, for which the environmental
effects on the EMRI waveform are  inevitable.  The possible degeneracy  calls for a systematic framework for searching new fundamental physics with EMRIs, with astrophysical environmental effects taken into account.

In the context of wet EMRIs, AGN jet physics and accretion physics are promising realms
where LISA and next-generation Event Horizon Telescope
(ngEHT \footnote{\url{https://www.ngeht.org/}}) may synergize.
According to the estimate in \cite{Pan:2020dze}, a fraction
of low-redshift $(z\lesssim 0.3)$ EMRIs can be traced back to their
host galaxies with LISA observations alone, and host AGNs of $\sim 50\%$
of low-redshift $(z\lesssim 0.5)$ wet EMRIs can be identified considering the
much lower density of AGNs.
Combining GW observations of wet EMRIs with radio obervations of AGN jets by, e.g. ngEHT, one can simultaneously measure the MBH mass $M_\bullet$, the MBH spin $\hat{\bm{a}}$,
the rotation direction of the accretion disk $\hat{\bm{L}}$, the jet power $\dot E_{\rm jet}$ and the jet direction $\hat{\bm{n}}_{\rm jet}$. This set of observables provide unprecedented opportunites to probe the AGN jet physics.
For example, an ensemble of events with $\{ \hat{\bm{n}}_{\rm jet} \cdot \hat{\bm{a}}, \hat{\bm{n}}_{\rm jet}  \cdot \hat{\bm{L}}\}$ data may help us to constrain various jet launching models, i.e., powered by the rotating energy of the MBH \cite{Blandford1977} or by the accretion disk \cite{Blandford:1982di}.
In addition, certain disk properties are directly constrained with GW observations
via the disk environmental effects on the EMRI waveform \cite{Yunes2011,Kocsis2011,Barausse2014},
and accretion physics of AGN disks is also one of the primary targets of ngEHT.

Wet EMRIs with host AGNs identified are ideal ``bright sirens" for constraining the late time cosmology
 (e.g., the Hubble constant and the equation of state of dark energy),
because  the  luminosity distance and the redshift can be measured from GW and electromagnetic observations,
respectively.  It will be interesting to compare the sensitivity of this method to other approaches,
with the predicted wet EMRI rate from this study.

Wet EMRIs encode additional information of MBH growth in their
orbital inclination angles $\iota_0$ with respective to the MBH spin.
If all MBHs grow up via coherent gas accretion where gas feeds are from a fixed direction,
orbital inclination angles of wet EMRIs at coalescence should be $\iota_0\approx \pi/2$.
If MBHs grow up via chaotic gas accretion from a random direction in each active phase,
a fraction of wet EMRIs form before the MBH spin direction $\hat{\bm{a}}$
is aligned with the disk rotation direction $\hat{\bm{L}}$
via the Bardeen-Petterson mechanism \cite{Bardeen1975},
and their orbital inclinations are approximately $\iota_0\approx \cos^{-1}(\hat{\bm{a}}\cdot \hat{\bm{L}})$.
In a similar way, MBH growth via different merger channels also imprints differently  on the
inclinations of wet EMRIs.

\acknowledgements
  We thank Hui-Min Fan and Alberto Sesana  for very helpful discussions.
  The authors are supported by the Natural Sciences and Engineering Research
  Council of Canada and in part by Perimeter Institute for Theoretical Physics.
  Research at Perimeter Institute is supported in
  part by the Government of Canada through the Department of
  Innovation, Science and Economic Development Canada and
  by the Province of Ontario through the Ministry of Colleges
  and Universities.

\bibliography{ms}

\end{document}